# Phase separation in iron chalcogenide superconductor Rb$_{0.8+x}$Fe$_{1.6+y}$Se$_2$ as seen by Raman light scattering and band structure calculations

Yu. Pashkevich[1], V. Gnezdilov[2], P. Lemmens[3], T. Shevtsova[1], A. Gusev[1], K. Lamonova[1], D. Wulferding[4], S. Gnatchenko[2], E. Pomjakushina[5], and K. Conder[5]

[1]*A.A. Galkin Donetsk Institute for Physics and Engineering, NAS of Ukraine, 03680 Kyiv, Ukraine*
E-mail: yu.pashkevich@gmail.com

[2]*B.I. Verkin Institute for Low Temperature Physics and Engeneering, NAS of Ukraine, 61103 Kharkov, Ukraine*

[3]*Institute for Condensed Matter Physics, TU Braunschweig, D-38106, Germany*

[4]*Center for Artificial Low Dimensional Electronic Systems, Institute for Basic Science, Pohang 790-784, Korea*

[5]*Laboratory for Developments and Methods, PSI, CH-5232 Villigen PSI, Switzerland*

We report Raman light scattering in the phase separated superconducting single crystal Rb$_{0.77}$Fe$_{1.61}$Se$_2$ with $T_c$ = 32 K over a wide temperature region 3–500 K. The observed phonon lines from the majority vacancy ordered Rb$_2$Fe$_4$Se$_5$ (245) antiferromagnetic phase with $T_N$ = 525 K demonstrate modest anomalies in the frequency, intensity and halfwidth at the superconductive phase transition. We identify phonon lines from the minority *compressed* Rb$_\delta$Fe$_2$Se$_2$ (122) conductive phase. The superconducting gap with $d_{x^2-y^2}$ symmetry has been detected in our spectra. In the range 0–600 cm$^{-1}$ we observe a weak but highly polarized $B_{1g}$-type background which becomes well-structured upon cooling. A possible magnetic or multiorbital origin of this background is discussed. We argue that the phase separation in M$_{0.8+x}$Fe$_{1.6+y}$Se$_2$ is of pure magnetic origin. It occurs below the Néel temperature when the magnetic moment of iron reaches a critical value. We state that there is a spacer between the majority 245 and minority 122 phases. Using *ab initio* spin-polarized band structure calculations we demonstrate that the *compressed* vacancy ordered Rb$_2$Fe$_4$Se$_5$ phase can be conductive and therefore may serve as a protective interface spacer between the purely metallic Rb$_\delta$Fe$_2$Se$_2$ phase and the insulating Rb$_2$Fe$_4$Se$_5$ phase providing percolative Josephson-junction like superconductivity all throughout of Rb$_{0.8+x}$Fe$_{1.6+y}$Se$_2$. Our lattice dynamics calculations show significant differences in the phonon spectra of the conductive and insulating Rb$_2$Fe$_4$Se$_5$ phases.

### Introduction

Analogous to oxygen intercalated La$_2$CuO$_{4+d}$ [1], an intrinsic structural phase separation into the antiferromagnetic insulating and nonmagnetic superconducting phases has been shown to occur in the alkali iron selenides M$_{0.8+x}$Fe$_{1.6+y}$Se$_2$ (M = K, Rb, Cs alkali metals) [2–14]. Immediately after this discovery the question about coexistence and competition of two phases arose. While one phase evidences robust superconductivity with $T_c$ ~ 30 K the other phase develops antiferromagnetic (AFM) order with an anomalously large Néel temperature of $T_N$ ~ 500 K and with the highest magnetic moment ~ 3.3 μ$_B$/Fe. Experimental data collected at an early stage favored a simple phase separation scenario with two structurally distinctive and non-interacting phases. However, following studies pointed out that some remnant interplay occurs [15,16]. Particularly, it was noticed that an intermediate phase should exist between insulating and superconducting regions [17] and that the AFM phase can be related to the unusual electronic properties of $M_{0.8+x}Fe_{1.6+y}Se_2$ compounds through formation of an orbital-selective Mott phase [18–21].

In this paper we examine spectral fingerprints of the nonmagnetic superconducting phase and the insulating antiferromagnetic phase in the superconductor $Rb_{0.77(2)}Fe_{1.61(3)}Se_2$ with $T_c = 32$ K using Raman light scattering and *ab initio* band structure and lattice dynamics calculations. So far, only one Raman study on $K_{0.68}Fe_{1.57}Se_2$ has considered a phase separation [22] whereas in other investigations the Raman spectra have been analyzed using the of the one phase concept [23–28]. In addition, the Raman studies of the rubidium based compound mainly focused on superconducting properties of the sample [28].

To date much information about details of the phase separation in the $M_{0.8+x}Fe_{1.6+y}Se_2$ (M–Fe–Se) compounds has been accumulated. General consensus has been achieved among the following observations: 1) The insulating phase is an antiferromagnetic semiconductor [8] with an enormously large Néel temperature ($T_N \sim 500$ K) an iron magnetic moment of 3.3 $\mu_B$ [29] which is highest among all iron superconductor parent compounds; 2) The antiferromagnetic phase represents an iron vacancy ordered layered structure $\sqrt{5} \times \sqrt{5} \times 1$ that corresponds to the so-called "245" stoichiometry with two formula units of the $M_2Fe_4Se_5$ in the primitive cell and with space group $I4/m$ [6,7]; 3) The phase separation sets in at a temperature $T_P$ that is a few tens of Kelvin below $T_N$ which in turn occurs about 10–20 K below the vacancy ordering temperature $T_S$ [4–6,13,30]; 4) The volume fraction ratio of metallic to antiferromagnetic phase is about 1/9, i.e., the majority constitutes of the insulating vacancy ordered phase [3,6,9,13], triggering a question about the percolative limit in the superconductive state; 5) The metallic nonmagnetic phase has an averaged $I4/mmm$ tetragonal symmetry with set of Bragg rods $\sqrt{2} \times \sqrt{2} \times 1$ which is expected for an averaged vacancy disordered structure [6,7,13,14]. The respective lattice constants are compressed in the $ab$ plane and expanded along $c$ axis compared to the AFM phase [4–7,13,14]. 6) The minority metallic phase forms a complex 3D nanoscale stripe-like network embedded into the insulating phase with plate-shaped features aligned along the {113} planes and elongated along the ⟨301⟩ directions [10,16,31,32]. Such a structure suggests that the superconductive state in $M_{0.8+x}Fe_{1.6+y}Se_2$ may be realized through Josephson junctions. The latter behavior has been observed in optical conductivity studies [2,33]. 7) In both phases the $c$ axes coincide which means both phases are crystallographically coherent [3–5,16].

The exact structure of the metallic superconductive phase is not known. Its average structure can be well described by the $I4/mmm$ $BaFe_2As_2$ structure but with alkali deficient chemical content $M_\delta Fe_2Se_2$ ($\delta < 1$). This model approximation was firstly suggested in the paper [13] and now this crystallographic attribution of the superconductive phase is generally accepted [34,35]. However, the minority phase $M_\delta Fe_2Se_2$ has never been successfully synthesized separate from the majority $M_2Fe_4Se_5$ phase. Recently [16] it was explained by specific mechanism of the superconductive phase formation which relies on the occurrence of an "imperfect" iron-vacancy order-disorder transition. The superconductive phase is a remnant of a high-temperature iron-vacancy-disordered phase with more iron concentration but less alkali ion concentration compared to the iron-vacancy-ordered phase.

In this scenario of phase separation the minority phase should consist mainly of the $M_\delta Fe_2Se_2$ vacancy free phase together with a small amount of phases which include iron vacancies. Using *ab initio* band structure calculations we demonstrate that the $M_2Fe_4Se_5$ vacancy ordered *compressed* phase, can be in the metallic state. We suggest that this third phase can serve as an interface to ensure the percolative superconductivity in $M_{0.8+x}Fe_{1.6+y}Se_2$. We provide lattice dynamics calculations for both metallic and insulating $Rb_2Fe_4Se_5$ phases to distinguish their phonon spectra. However, in spite of a large difference observed in the high-frequency part of our spectra we did not find a direct manifestation of this third phase in our Raman spectra.

The distinctive feature of the $M_{0.8+x}Fe_{1.6+y}Se_2$ superconductors is the absence of hole pockets near the Brillouin zone (BZ) center that were established by angle resolved photoemission spectroscopy studies [12,36–40]. This discovery has altered the dominant channel of superconductive pairing from the $s_+$-type, which is based on the interaction between electron and hole pockets, to the $d_{x^2-y^2}$ channel which based on the interaction between electron pockets on the BZ boundary [41]. The $d_{x^2-y^2}$ channel has been detected in Raman studies of $Rb_{0.8}Fe_{1.6}Se_2$ [28]. However, another important feature of $M_{0.8+x}Fe_{1.6+y}Se_2$ compounds is an insensitivity of the superconducting transition temperature to the doping of alkali ions $\delta$ in the metallic phase $M_\delta Fe_2Se_2$ which remains at $T_c \sim 30$ K for a wide doping range [42,43]. The alkali content $\delta$ determines the electron doping level and highly affects the topology of Fermi surface [44]. Therefore one should expect that the structure of the superconductive gap in $M_{0.8+x}Fe_{1.6+y}Se_2$ should display some modifications which depend on the history of sample preparation (e.g., it should be a function of the doping level $\delta$). In our Raman spectra we observed $d_{x^2-y^2}$ symmetry of the superconductive gap, typical for iron selenides while some specific features are different from the Raman spectra of $Rb_{0.8+x}Fe_{1.6+y}Se_2$ reported previously [28]. Additionally our first-principal band structure calculations for the metallic $RbFe_2Se_2$ modeling phase reveal an absence of hole pockets at the center of the Brillouin zone.

We did not see a noticeable change in the volume of superconductive fraction upon decreasing temperature through $T_c$ in contrast to the phase separated $La_2CuO_{4+d}$ [45]. However, our observations reveal some alteration of the electronic structure in the metallic phase which starts slightly above $T_c$.

## Experimental details

A $Rb_{0.8+x}Fe_{1.6+y}Se_2$ single crystal with chemical content $Rb_{0.77(2)}Fe_{1.61(3)}Se_2$ was synthesized following the procedure described in [46]. Chemical homogeneity and elemental composition of the cleaved crystals were studied using x-ray fluorescence spectroscopy (XRF, Orbis Micro-XRF Analyzer, EDAX). Elemental distribution maps for Rb, Fe and Se were collected in a vacuum applying white x-ray radiation produced by a Rh tube (35 kV and 500 μA). The x-ray primary beam was focused to a spot of 30 μm diameter. A primary beam Ti filter (25 μm thickness) was applied. An area of ~ 0.5 cm$^2$ was scanned. Prior to the measurements, elemental calibration was done using as a standard carefully weighted, homogenized, and pressed into a pellet mixture of Se, Fe, and corresponding alkali metal carbonates. The applied calibration procedure results in ~ 2% accuracy of the determined stoichiometric coefficient values. The superconducting properties of the plate-like crystal were characterized with a Quantum Design MPMS XL SQUID magnetometer. The temperature dependence of the magnetic susceptibility is shown in Fig. 1.

The crystal was cleaved and a flat shiny surface was obtained. The freshly-cleaved sample was immediately transferred into an evacuated cryostat for Raman scattering (RS) studies avoiding surface contamination and degradation. Raman scattering measurements were performed in quasibackscattering geometry with the excitation line $\lambda = 532.1$ nm of a solid-state laser. Raman spectra were measured from the crystallographic $ab$ plane. The laser power of less than 3 mW was focused to a 0.1 mm diameter spot on the sample surface. Spectra of the scattered radiation were collected by a Dilor-XY triple spectrometer and a micro-Raman setup (Horiba Labram) equipped with liquid nitrogen cooled charge coupled device (CCD) detector (Horiba Jobin-Yvon, Spectrum One CCD-3000V). Temperature dependences were obtained in a variable temperature closed cycle cryostat (Oxford/Cryomech Optistat, RT-2.8 K). We have collected additional Raman spectra over an extended temperature range to 500 K in a constantly evacuated heating chamber.

## Results and discussion

The tetragonal insulating majority phase of $Rb_{0.8+x}Fe_{1.6+y}Se_2$ possesses a $\sqrt{5} \times \sqrt{5} \times 1$ iron vacancies ordering pattern that is described by $I4/m$ ($N$ 87) space group symmetry with two $M_2Fe_4Se_5$ formula units per primitive cell (so-called "245" phase). The Fe ions occupy general type 8i $(x,y,z)$-positions, while the Se ions are distributed between 8i $(x,y,z)$ (Se1)- and 2e $(1/2,1/2,z)$ (Se2)-positions, and the Rb ions reside at 4h $(x,y,0)$-positions. Note that Se2 atoms are located at special positions in the layer of $FeSe_4$ tetrahedrons, being the corner shared ions between four tetrahedrons in the same plaquet while the connection between different plaquets is realized through Se1 atoms. In spite of the high symmetry of Se2 ions their $z$ coordinate is not fixed. Thus, the free coordinates of the Se1 and the Se2 sites make the structure of the $M_2Fe_4Se_5$ compounds much more susceptible to possible orbital re-ordering processes and iron spin state changes compared to other chalcogenides, like FeSe and FeTe. The highly distorted $FeSe_4$ tetrahedron shape, in which the Fe–Se2 distance is substantially larger than the Fe–Se1 one, also supports this tendency.

The block checker-board type of antiferromagnetic order in $Rb_{0.8+x}Fe_{1.6+y}Se_2$ is formed by a plaquet (square) of four nearest-neighbor (in block) iron atoms with ferromagnetic order along $c$ axis while other in-plane and out-of-plane nearest-neighbor plaquets are ordered antiferromagnetically [6,7,9]. There is no multiplication of the crystallographic primitive cell. The neutron diffraction studies as well polarized muon spin relaxation measurements did not reveal the spin reorientation phase transition in this compound under lowering temperature [7,9]. This type of magnetic order parameter (belonging to the $\tau_2$ or $A_u$ irreducible representation [6]) breaks inversion symmetry but retains tetragonal symmetry. The unitary subgroup of magnetic group $I4/m'$ is $I4$ ($N$ 79) with rotational symmetry $C_4$ so that the magnetically ordered part of the sample remains in the ferroelectric state at all temperatures of our experiment. The crystallographic and magnetic structures of $Rb_2Fe_4Se_5$ are shown in Fig. 2.

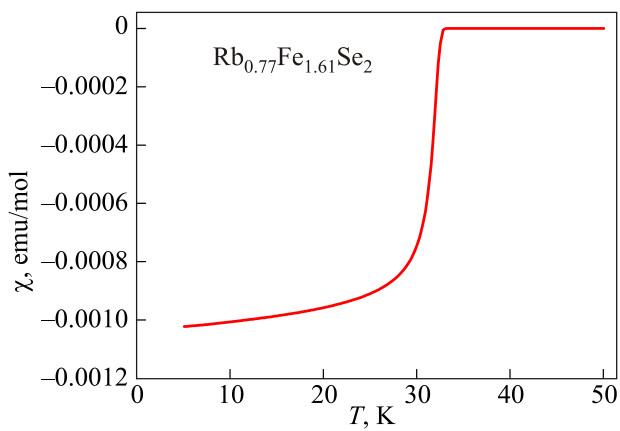

*Fig. 1.* Temperature dependence of the magnetic susceptibility of our $Rb_{0.77}Fe_{1.61}Se_2$ sample. The sharp transition into the superconductive state at $T_c = 32$ K demonstrates the high quality of the crystal.

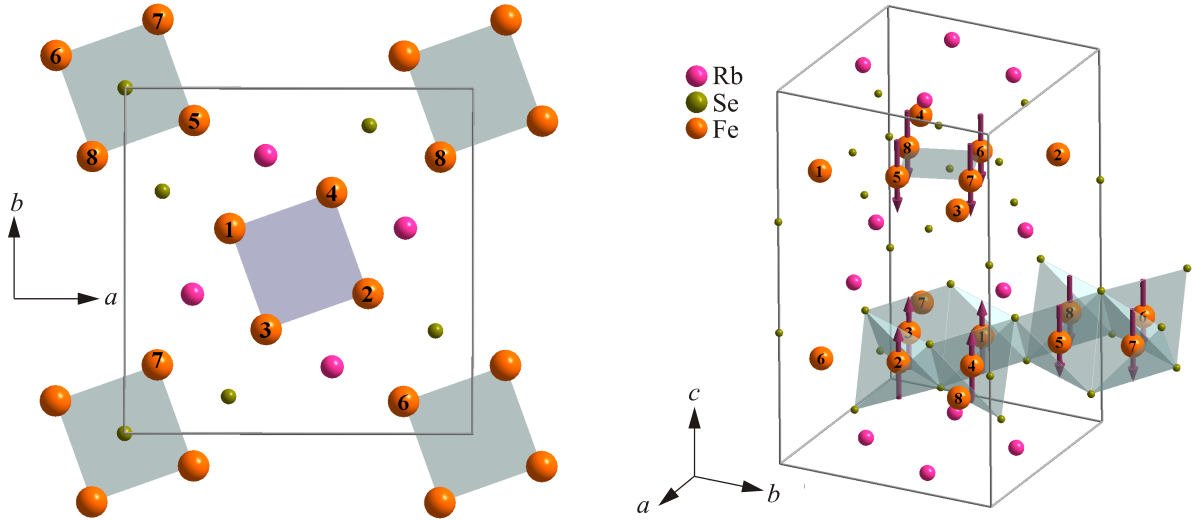

*Fig. 2.* (Color online) The crystallographic and magnetic structures of Rb$_2$Fe$_4$Se$_5$. The plaquets (blocks) of four nearest-neighbor iron atoms are marked by shaded planes on left side. The block type of AFM order with a layer of FeSe$_4$ tetrahedrons is shown on the right side.

The symmetry analysis reveals that for the M$_2$Fe$_4$Se$_5$ vacancy ordered structure in the paramagnetic phase there are in general 66 phonon modes including 3 acoustic ones. The mechanical representation Γ consists of the following irreducible representations (IRp):

$$\Gamma = 9A_g + 8B_g + 8E_g + 8A_u + 7B_u + 9E_u.$$

Out of 25 Raman-active modes only 17 modes ($9A_g + 8B_g$) are accessible for the given backscattering geometry of our Raman experiment with allowed incident-scattered light polarization for $A_g$(XX+YY, ZZ) modes and $B_g$(XX–YY, XY) modes. Here the X and Y axes are directed along the a and b lattice constants of the 245 phase shown in Fig. 2. Raman-active modes are distributed among the atom positions in following way: 4h (Rb) = $2A_g + 2B_g$; 8i (Se1 and Fe) = $3A_g + 3B_g$; and 2e (Se2) = $A_g$.

In accordance with the generally accepted description the minority phase is alkali deficient, iron vacancy free, nonmagnetic and metallic. It consists of layers of the edge shared FeSe$_4$ tetrahedrons stacked along the c axis with the Rb ions in between. This Rb$_\delta$Fe$_2$Se$_2$ phase (so-called "122" phase) has an averaged *I*4/*mmm* symmetry [13] with Rb content δ depending on the initial x, y chemical composition of the Rb$_{0.8+x}$Fe$_{1.6+y}$Se$_2$ [42,43]. The occupancy of atoms (Fe at 4d, Se at 4e, and Rb at 2a) should result in two additional Raman active modes $A_{1g}$(Se) and $B_{1g}$(Fe) into Raman spectra measured in our geometry of scattering. The Raman tensor of the $B_{1g}$(Fe) mode has (X′X′–Y′Y′) symmetry. Here the X′ and Y′ axes of the 122 phase are rotated by 27° from the X–Y axes of the 245 phase. The alignment of the $C_4$ axes remains the same for the 245 and the 122 phases [3–5].

The evolution of the Rb$_{0.77}$Fe$_{1.61}$Se$_2$ Raman spectra in parallel (XX) and crossed (YX) polarizations as a function of temperature from 3 to 490 K is shown in Fig. 3. In total at least fifteen phonon modes can be clearly identified from both polarizations. All of these phonon modes are located in the frequency region below 300 cm$^{-1}$ similar to previous studies of K$_{0.8+x}$Fe$_{1.6+y}$Se$_2$ [24,25,27]. The number of observed lines evidences the presence of the vacancy ordered state already at T = 490 K just below the Néel temperature $T_N$ = 525 K.

The tetragonal symmetry of the sample is clearly seen in the polarization dependence of the spectra. This is illustrated in Fig. 4 where we compare the low-temperature XX and YX spectra. The $A_g$ and $B_g$ phonons can be easy distinguished in our well polarized spectra. However, it is not so straightforward to identify the $B_{1g}$ phonon of the minority *I*4/*mmm* phase from the $B_g$ phonons with strong intensity of the *I*4/*m* majority phase. In Raman studies of the K$_{0.8+x}$Fe$_{1.6+y}$Se$_2$ the weak $B_{1g}$ line of the 122 phase has been extracted by using its specific intensity dependence from the angle between polarizations of incident and scattered light [24], while the $A_{1g}$ phonon has been assigned under the assumption about of similar Raman spectra from (Sr,K)Fe$_2$As$_2$ and K$_\delta$Fe$_2$Se$_2$. The observed frequencies of the 122 phase at 85 K $A_{1g}$ (180 cm$^{-1}$) and $B_{1g}$ (207 cm$^{-1}$) [24] turn out to be strikingly equal (with the accuracy of one wave number) to the frequencies of the same symmetry phonons in FeSe at 100 K [47]. Relying on that identity we compare the Raman spectra of the FeSe [47] and Rb$_{0.77}$Fe$_{1.61}$Se$_2$ in Fig. 4. We did not observe a direct coincidence of the phonon lines in both spectra as it takes place in K$_{0.8+x}$Fe$_{1.6+y}$Se$_2$ [24]. However, considering the similarity of some lines in the spectra of both compounds, one can preliminary attribute the mode at 178.4 cm$^{-1}$ in the XX spectra as the $A_{1g}$ mode and the mode at 212 cm$^{-1}$ in the XY spectra at the $B_{1g}$ mode of the Rb$_\delta$Fe$_2$Se$_2$ metallic minority phase. Below we will give more evidence to substantiate our attribution.

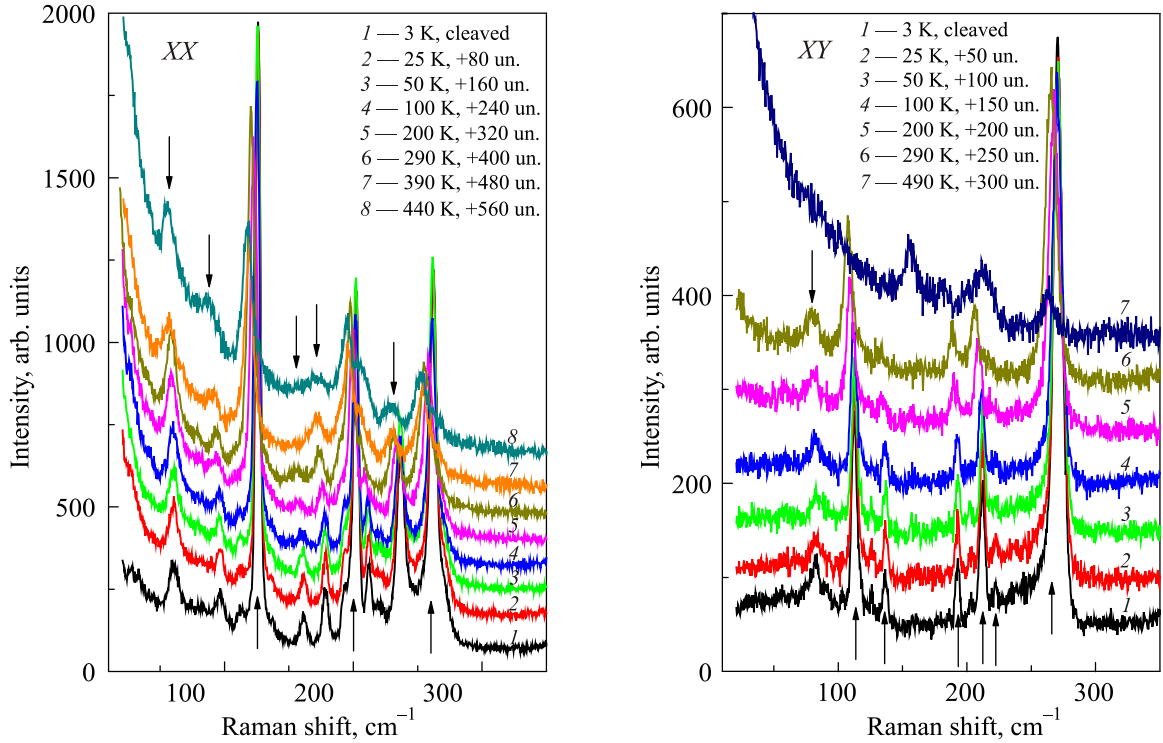

*Fig. 3.* Temperature dependences of $Rb_{0.77}Fe_{1.61}Se_2$ Raman spectra in the diagonal polarization for $A_g$-type phonon lines (left side) and crossed polarization for $B_g$-type phonon lines (right side). The observed phonons are marked by arrows.

The temperature-dependent Raman spectra of $Rb_{0.77}Fe_{1.61}Se_2$ phonon lines were described explicitly by Lorentz functions. All of them show a conventional phonon line shape without any evidence of the Fano-like asymmetry. In Figs. 5 and 6 the results of a temperature analysis of the representative phonons is shown. Interestingly, a like asymmetry indicative for an electron-phonon coupling did appear in the Raman spectra of the superconductive $K_{0.75}Fe_{1.75}Se_2$ [27], while being absent in the Raman spectra of the superconductive $K_{0.68}Fe_{1.57}Se_2$ [24]. Hence, the question arises whether there is a connection to details of the phase separation caused by different potassium content, which might change the doping level of the metallic phase as well as the ratio between minority and majority phase, whether this is simply an issue of sample quality.

We did not observe a structural phase transition which takes place in $K_{0.75}Fe_{1.75}Se_2$ at 250 K [27]. Instead, the frequencies and halfwidths of selected $A_g$ phonon lines show a small kink around 270 K in their temperature dependences shown in Fig. 5. Among high-frequency $B_g$ phonons with iron-dominated vibrations the $B_g(5)$ phonon shows the strongest frequency increase under cooling. This resembles a frequency anomaly of the $B_{1g}$ phonon in FeSe detected in our Raman studies [47], supporting our assignment of the $B_g(5)$ phonon as $B_{1g}$ phonon of the 122 phase with $I4/mmm$ symmetry.

Further evidence is found in the behavior of the scattering intensity, which increases anomalously for the $B_g(5)$ phonon under approaching and just below superconductivity onset, illustrated in Fig. 7. The main contribution to the intensity gain comes from the $B_{1g}(XX–YY)$ channel that is clearly seen from comparing of the Raman spectra in crossed and RL-polarizations (Fig. 7, right panel). Here and below the RL (RR) denotes opposite (the same) chirality of the incident and scattered light with right (R) or left (L) circular polarizations. Simultaneously some inelastic wide band bump with distinct $B_{1g}(XX–YY)$ symmetry continually increases under cooling and persists in the superconductive state.

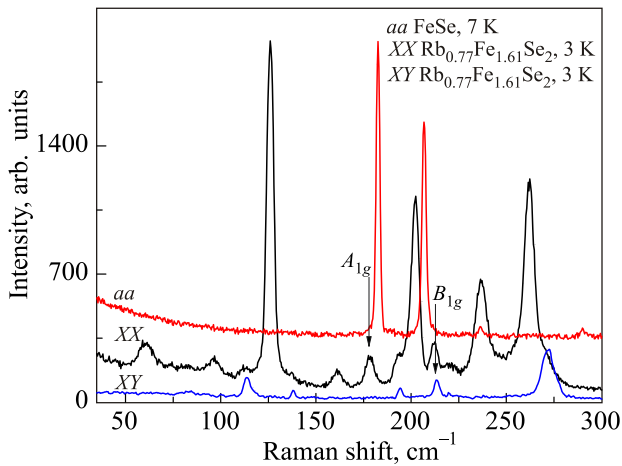

*Fig. 4.* Comparison of the *XX* and *XY* Raman scattered polarized spectra of $Rb_{0.77}Fe_{1.61}Se_2$ at 3 K with the *aa*-polarized spectra of FeSe at 7 K adopted from [47]. All strong $A_g$ lines from the *XX* polarization are absent in the *XY* polarization. Almost all $B_g$ lines which are seen in the *XY* polarization have visible counterparts in the *XX* polarization that is in accordance with $I4/m$ symmetry of the $B_g$ Raman tensor (*XX–YY*, *XY*). The one exception is the highest and most intensive $B_g$ mode which has a very weak counterpart in the *XX* spectra. The assignment of the $A_{1g}$ and $B_{1g}$ phonon lines to the minority $I4/mmm$ metallic $Rb_\delta Fe_2Se_2$ (122) phase are marked by arrows.

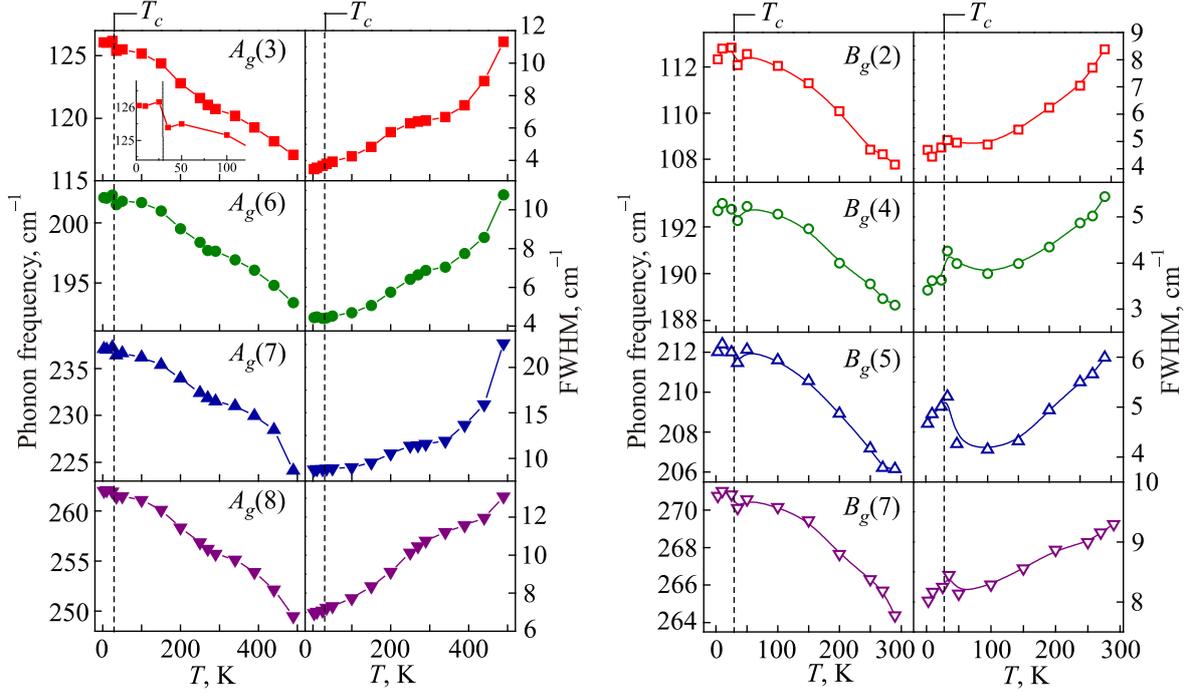

*Fig. 5.* Temperature dependent parameters of selected $A_g$ (left panel, solid symbols) and $B_g$ (right panel, open symbols) phonon modes in $Rb_{0.77}Fe_{1.61}Se_2$. Temperature dependences of the frequency (left side of each panel) and linewidth, FWHM (right side of each panel). Solid lines are guides to the eyes.

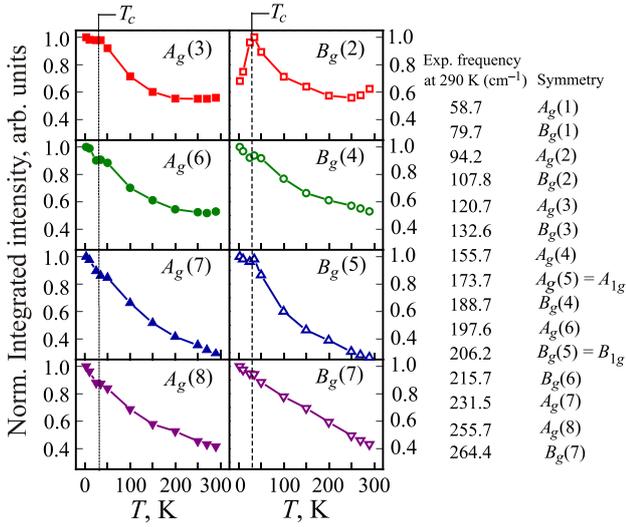

*Fig. 6.* Temperature dependences of the intensity of selected $A_g$ (left panel, solid symbols) and $B_g$ (right panel, open symbols) phonon modes in $Rb_{0.77}Fe_{1.61}Se_2$. Solid lines are guides to the eyes. On the right side we display the symmetry assignment, enumeration and frequency of observed phonons.

In spite of the same symmetry of the broad band and the $B_{1g}$ phonon at 213 cm$^{-1}$ there is no visible Fano-like spectral feature. The intensities of the bump band and the $B_{1g}$ phonon are both nicely fitted by an oscillator-like shape and a Lorentz function, respectively. Similarly, the $B_g(7)$ phonon at 272 cm$^{-1}$ also does not show Fano-like features. However, in the last case there is no $XX–YY$ contribution to the Raman tensor of the $B_g(7)$ phonon. This is supported by the similarity in the temperature evolution of the $B_g(7)$ intensity in both RL and crossed polarizations (Fig. 7) as well by the absence of significant contribution of $B_g(7)$ phonon mode in the $XX$ — Raman spectra (Fig. 4). In other words the $B_g(7)$ $XY$ phonon cannot interact with excitations manifested in the $XX–YY$ bump. The absence of a Fano-like features for the $B_{1g}$ phonon at 212 cm$^{-1}$ can be interpreted in two ways: (i) the bump excitations originate from another phase, e.g., from the insulating 245 phase and cannot interact with a phonon of the 122 metallic phase, regardless of the symmetry. (ii) the bump band does not present a many-particle continuum excitation but rather a set of one-particle excitations and therefore no Fano-like interaction is expected. Below we will discuss both possibilities.

The anomalous intensity increase of the $B_{1g}$ phonon is not related to a possible increase of the 122 phase volume fraction as we do not see the simultaneous increase of the $A_{1g}$ phonon intensity. At the same time the $B_{1g}$ mode does not evidence any frequency shift, thereby signaling a highly specific electron structure alteration as the origin for the intensity anomaly that contributes solely to the $B_{1g}$-type

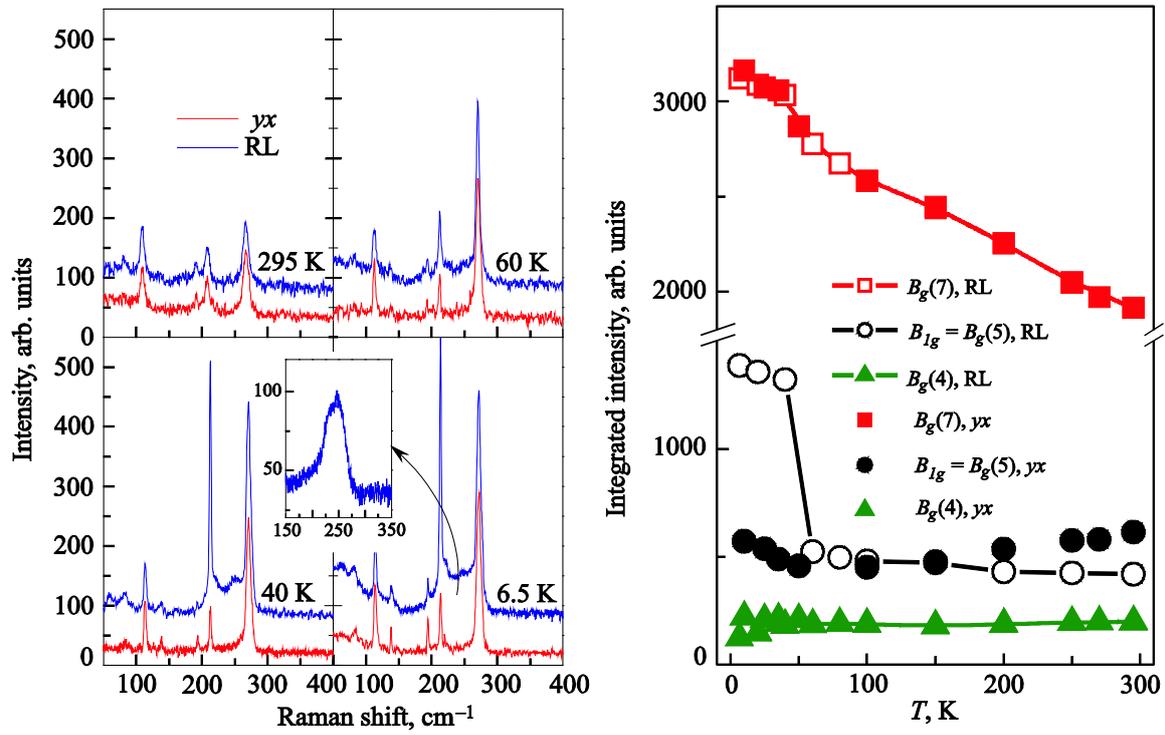

*Fig. 7.* (Color online) Comparison of Raman spectra for crossed and RL polarizations at different temperatures (left panel). Temperature-dependent intensities of selected $B_g$ phonons for different polarizations (right panel). The inset the left panel shows a broad band of $B_{1g}(XX–YY)$ symmetry which is centered around 245 cm$^{-1}$.

Raman tensor and not to the $A_{1g}$-type Raman tensor. The difference in the electronic structure above and below $T_c$ in Rb 245 has been detected in ARPES studies [12]. Note that we observe the change in intensity already at $T = 40$ K just before the onset superconductivity.

The above mentioned bump belongs to the low intensive background which we obtain from the Raman spectra by subtracting the phonon contribution. The background becomes clearly structured under cooling and persists in the superconductive state (see Fig. 8). This background can be described very well with four wide oscillator shaped bands.

It has essentially $B_{1g}(XX–YY)$ symmetry in which the nondiagonal $XY$ components are absent. The background almost disappears in the nondiagonal spectra by respective choice of the axes (for instance it shows up in the $YX$ spectra shown in the Fig. 3 but it is absent in the $yx$ spectra shown in the Fig. 7).

Let us discuss the possible origin of this background. The block type AFM order with eight magnetic ions in the primitive cell should lead to four double degenerated transversal spin waves at $k = 0$. The energy of the first wide band of background equals to 50 cm$^{-1}$. It nicely coin-cides with a gap of the acoustic transversal spin wave that has been found below 8 meV by inelastic neutron scattering in the superconductive $Rb_{0.82}Fe_{1.68}Se_2$ with $T_c = 32$ K [48] as well with spin wave gap at 6.5 meV found in the superconductive $(Tl,Rb)_2Fe_4Se_5$ with $T_c \sim 32$ K [49]. Due to the specific symmetry of acoustic spin waves this possible one-magnon scattering should not appear in the $B_{1g}$ channel neither by Faraday nor by exchange mechanism scattering. Further bumps are centered at 247, 355 and 485 cm$^{-1}$ and have no counterparts in the optical spin wave spectra at $k = 0$ ($\Gamma$ point of Brillouin zone) [48]. However, a respective symmetry analysis shows that one-magnon scattering in the $B_{1g}$ channel is possible on an optical branch of the longitudinal magnons. It is caused by the exchange mechanism of scattering. Such a scattering can be assisted by specific phonons with vibration of atoms that are involved into pathway of the superexchange interactions. Interestingly the frequency of the most prominent background bump at 247 cm$^{-1}$ lies exactly in between frequencies of the two highly intensive $A_g$ phonons. One can demonstrate that this phonon type can assist longitudinal iron moment oscillations.

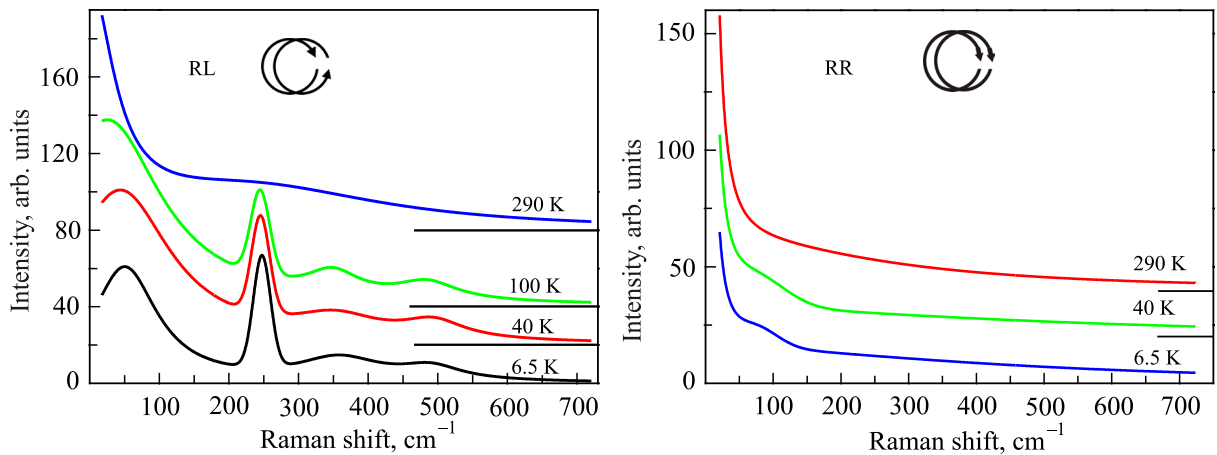

*Fig. 8.* (Color online) Temperature dependences of the low-energy electronic background for two different polarizations in Rb$_{0.77}$Fe$_{1.61}$Se$_2$. The smooth development of the well defined structure is clearly seen under decreasing temperature in the RL polarization. This structure persists even in the superconductive state.

Another possible explanation of the structured and polarized background can be related to the multiorbital nature of the iron HTSC. The multiband electronic structure allows some interband transitions across the Fermi surface caused by charge fluctuations [50,51] and a $d$-wave Pomeranchuk instability [52]. According to Ref. 50 the scattering in the $B_{1g}$-symmetry channel corresponds to intraorbital transitions at the electron pockets, which are expected in our 122 metallic phase. The energy scale of the background bumps also points towards a metallic phase origin. Below we will see that the electronic band structure of the metallic *compressed* $Rb_2Fe_4Se_5$ phase allows some low-energy interband transitions. If we assume this mechanism to be valid in the insulating 245 phase then the energy of the respective interband excitations should exceed the energy of the semiconducting gap that is roughly 0.4 eV. Interestingly, the highly polarized broad bump, centered at low $T$ at 450 cm$^{-1}$, has also been observed in FeSe [47].

Raman spectroscopy is a suitable tool to elucidate the symmetry of the superconducting gap [53]. The distinctive difference between superconducting and normal states appears in RL polarized spectra but is absent in spectra of RR polarization (see insets in Fig. 9). This difference evidences the presence of at least two types of superconducting gaps with $B_{1g}$ and $B_{2g}$ symmetries in our $Rb_{0.77}Fe_{1.61}Se_2$ sample. As shown in [54] the gap with pure $d_{x^2-y^2}$ symmetry should be the dominant superconducting gap in the iron selenides where the Fermi surface of the metallic phase contains electronic pockets at the BZ boundary and does not have the hole pockets at the center of BZ. The presence of a subdominant $B_{2g}$ superconducting gap should be seen in Raman spectra as the appearance of new low-energy excitations in the superconducting state which could be interpreted as Bardasis–Schrieffer modes [55]. In iron-based superconductors these collective modes can arise due to a coexistence and competition of two different pairing channels [56]. Since we did not see new features in the superconducting state we conclude that the $B_{2g}$ superconducting gap is absent in $Rb_{0.77}Fe_{1.61}Se_2$. In fact, a previous study reported the absence of a $B_{2g}$ superconducting gap in $Rb_{0.8}Fe_{1.6}Se_2$ as well [28]. On other hand, collective in-gap modes have been found in $K_{0.75}Fe_{1.75}Se_2$ in the RL channel [27] which evidences a coexistence and strong competition of the $s$-wave and the $d$-wave pairing mechanisms [57]. The discrepancy in details of the superconducting state between Rb- and K- relatives of the $M_{0.8+x}Fe_{1.6+y}Se_2$ superconductors is connected with an uncontrolled and different electron doping value $\delta$ in the metallic phase $M_\delta Fe_2Se_2$ for the samples of different origin. Such an uncertainty is reflected in the phase diagram of $M_{0.8+x}Fe_{1.6+y}Se_2$ where the samples of different alkali metal and iron content nevertheless fall within the superconducting dome [42,43]. The relation between different pairing channels and the level of doping in M$_\delta$Fe$_2$Se$_2$ has been analyzed in a recent paper [58]. To this point one can add a difference between Raman spectra of the superconducting phase in Rb$_{0.8+x}$Fe$_{1.6+y}$Se$_2$ samples taken from different sources. While our spectra of Rb$_{0.77}$Fe$_{1.61}$Se$_2$ did not show any intensity gain in superconducting state for the phonons of insulating 245 phase at 80 and 115 cm$^{-1}$ spectra of Rb$_{0.8}$Fe$_{1.6}$Se$_2$ clearly evidence changes in $B_{1g}$ intensity [28].

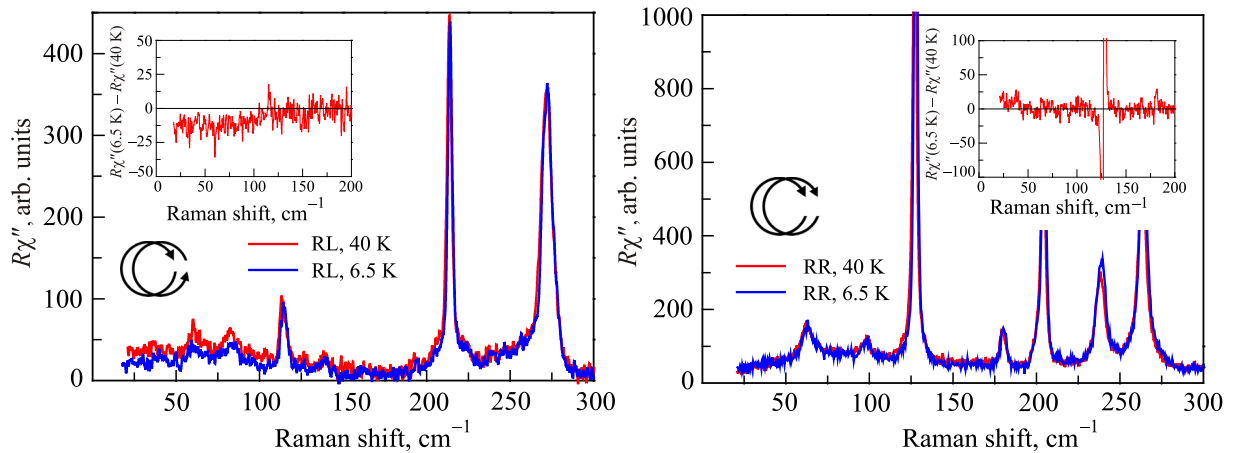

*Fig. 9.* (Color online) Manifestation of the superconducting state in Raman spectra of $Rb_{0.77}Fe_{1.61}Se_2$ for the RL (left) and RR (right) circular polarizations. The spectra are corrected by Bose factor. The insets plot the intensity difference of the spectra above and below $T_c = 32$ K.

We did not see additional modes caused by the loss of inversion symmetry in the magnetically ordered and ferroelectric 245 phase. In this phase the 15 far-infrared active modes ($7A_u + 8E_u$) as well 7 previously silent $B_u$ modes should be Raman active. The group–subgroup relation between the paramagnetic phase (belonging to the $-I4/m$ group) and the unitary subgroup ($-I4$) of magnetic group of the antiferromagnetic phase yields the following transformations: $A_g$ and $A_u$ transition to $A$ ($XX+YY$, $ZZ$); $B_g$ and $B_u$ transition to $B$ ($XX-YY$, $XY$); and $E_g$ and $E_u$ transition to $E$ ($XZ$, $YZ$). In particular, $14$ additional modes (previously of $7A_u + 7B_u$ symmetry) should be seen in our scattering geometry. However, our *ab initio* band structure and lattice dynamic calculations demonstrate that the mixing between former $g$- and $u$-type of modes in the magnetically ordered phase is very small and can be neglected. This is due to an almost indiscernibly difference between the $I4/m$ and the $I4$ crystallographic structures which has been obtained under structure optimization. Hence we conclude that $u$-type modes of the Rb$_2$Fe$_4$Se$_5$ insulating magnetically ordered phase can not be observed in Raman experiment.

Finally, our phonon Raman spectra which can be assigned to the insulating vacancy and antiferromagnetically ordered Rb$_2$Fe$_4$Se$_5$ phase show several distinctive features at $T_c$ (see Figs. 5 and 6): 1) clear frequency jump of the $A_g$ modes at $T_c$; 2) frequency and linewidth anomaly of the $B_g$ modes at $T_c$; 3) and an intensity anomaly at $T_c$.

The observed anomalies are reminiscent of the superconductivity-induced features in the μSR spectra of FeSe under pressure [59] and in optical conductivity spectra of Rb [8] and K [60] members of the "245" family. It can be explained by a rearrangement in the electronic subsystem trough under the onset of superconductivity. A remarkable alteration of the low-frequency range (below 8 meV) in the optical response with a kink at $T_c$ has been observed in Rb$_2$Fe$_4$Se$_5$ [8]. Furthermore, in this frequency region they observed a double peak structure around 2 meV appears which is buried by the electronic background at high temperatures but persists into the superconducting state. Particularly interesting is a drop of the μSR frequency below $T_c$ in FeSe in the regime where magnetic and superconductive states coexist [59]. This drop signals the lowering of the absolute iron magnetic moment. The change in magnetic moment, in turn, reflects the change of electron density distribution among $d$-orbitals in the magnetically ordered phase, which nevertheless occurs below the superconducting phase transition.

**First-principal band structure and lattice dynamic calculations**

Density-functional (DFT) calculations can provide an adequate description of the band structure and the lattice dynamics of iron pnictides and chalcogenides if the iron spin states are explicitly taken into account [61–64]. We applied the all-electron full-potential linearized augmented — plane-wave method (ELK code) [65] with revised local spin density approximation [66] for the exchange-correlation potential in the spin-polarized mode. To obtain band structure and phonon frequencies for the given lattice parameters and magnetic structure without the full optimization procedure we assume that at the true value of magnetic moments the crystal structure is already optimized. Thus, for the first step of calculations we use the value of magnetic moment as a fitting parameter to optimize the ion's coordinates while experimental lattice constants remain to be fixed. We successfully applied this procedure in previous lattice dynamic calculations of the FeTe [67,68] and FeSe [47].

We used the experimental unit-cell parameters taken at room temperature for Rb$_{0.8+x}$Fe$_{1.6+y}$Se$_2$ [13] that adopt the vacancy ordered structure and $I4/m$ space symmetry with two formula units in the primitive cell with iron atoms located at the general 8(i) positions, Rb atoms at 4(h) positions, and Se atoms at two positions 8(i) and 2(e). For the self-consistent and phonon calculations we considered a 5×5×5 gamma-centered k-mesh, which corresponds to 21 points in the irreducible part of the Brillouin zone. The phonon frequencies were calculated in the "fix-spin" mode at which the iron magnetic moment was fixed at 2.62 μ$_B$/Fe and does not change throughout the calculations. The magnitude of the iron magnetic moment was obtained from our preliminary calculations which were carried out for the antiferromagnetic state of the Rb$_2$Fe$_4$Se$_5$ with fixed lattice constants. The magnetic structure has been accounted for explicitly by the so-called "block antiferromagnetic order" in which a plaquet of the four nearest-neighbor iron atoms has magnetic moments aligned ferromagnetically along the $c$ axis and all nearest-neighbor plaquets align antiferromagnetically (see Fig. 2). The calculated magnitude of the iron magnetic moment is in good agreement with experimental data taken at 300 K [13]. The partially optimized structural data are summarized in the Table 1.

Our band structure calculations for the vacancy ordered magnetic state result in the insulating solution shown in Fig. 10 for the stoichiometric Rb$_2$Fe$_4$Se$_5$ compound similar to the results mentioned in [8,12,69,70] for the whole stoichiometric "245" family. The density of states (DOS), presented in Fig. 11, highlights that this insulating state can be qualified as a semiconducting state with a very small interband gap of around 0.2–0.3 eV that is in accordance with experimental measurements of optical conductivity [8].

Table 1. The partially optimized structural data

| Parameter | Atomic and cell parameters (after optimization) | | | | | |
|---|---|---|---|---|---|---|
| | for the insulating phase of $Rb_2Fe_4Se_5$ | | | for the conductive phase of $Rb_2Fe_4Se_5$ | | |
| $M$, $\mu_B$/Fe | 2.62 | | | 0.01 | | |
| $a$, Å | 8.7996 | | | 8.5682 | | |
| $b$, Å | 8.7996 | | | 8.5682 | | |
| $c$, Å | 14.5762 | | | 14.636 | | |
| Position of atom | $x/a$ | $y/b$ | $z/c$ | $x/a$ | $y/b$ | $z/c$ |
| Rb (8h) | 0.399 | 0.803 | 0 | 0.3567 | 0.8122 | 0 |
| Fe (16i) | 0.2958 | 0.5920 | 0.2524 | 0.3061 | 0.6040 | 0.2489 |
| Se1 (16i) | 0.3880 | 0.7944 | 0.6495 | 0.4036 | 0.8022 | 0.6714 |
| Se2 (4e) | 0.5 | 0.5 | 0.1455 | 0.5 | 0.5 | 0.1596 |
| | Fe–Fe, Å | | Fe–Se, Å | Fe–Fe, Å | | Fe–Se, Å |
| | in block | between blocks | | in block | between blocks | |
| | 2.7872 | 2.896 | 2.4284 | 2.6636 | 2.6784 | 2.2718 |
| | | | 2.4238 | | | 2.2214 |
| | | | 2.4217 | | | 2.2717 |
| | | | 2.5124(Se2) | | | 2.2930(Se2) |

We carried out self-consistent band-structure calculations for the metallic vacancy free $RbFe_2Se_2$ phase (see Fig. 12). We used the structural parameters of the $I4/mmm$ phase (see Table 2) which has been identified at 200 K in the superconducting sample of $Rb_{0.8+x}Fe_{1.6+y}Se_2$ as the minority phase [35]. For the calculations we considered (2a) Rb site to be fully occupied. Calculations were carried out using the same parameters as previously, except for a change of the k-mesh into 9×9×5 due to the reduced size of the $RbFe_2Se_2$ unit cell compare to $Rb_2Fe_4Se_5$. The magnetic moment of Fe after self-consistent calculations without fixation turned out about 0 $\mu_B$. Our results demonstrate the absence of the hole-like Fermi surface at the BZ center and the presence of the electron-like double degenerated pockets at the BZ M-points that are in accordance with many angle-resolved photoemission experiments on alkali iron selenides.

We argue that the phase separation phenomenon in $M_{0.8+x}Fe_{1.6+y}Se_2$ is of magnetic origin. The driving force of the phase separation is an internal *in-plane* pressure, which develops *in the vacancy ordered phase* at temperatures below its AFM ordering temperature. The phase separation occurs when the internal pressure reaches a critical value under the static magnetic moment growth. Particularly in $Rb_{0.8+x}Fe_{1.6+y}Se_2$, the phase separation temperature coin-cides with the temperature at which the iron magnetic moments attain their maximum (see Fig. 8 in Ref. 13). Hence, the magnetic moment value is a key parameter which defines the level of internal nonuniform compression and the appearance of the vacancy free 122 phase. In the case of nonfixed boundaries of the sample the phase separation phenomenon is a response of the crystal to minimize the elastic free energy. Note that the influence of the suggested internal pressure is different from the influence of an external hydrostatic pressure. The latter should decrease all lattice constants whereas internal pressure has a nonuniform effect and leads to a decrease in $ab$ constants and an increase of the $c$ lattice constant.

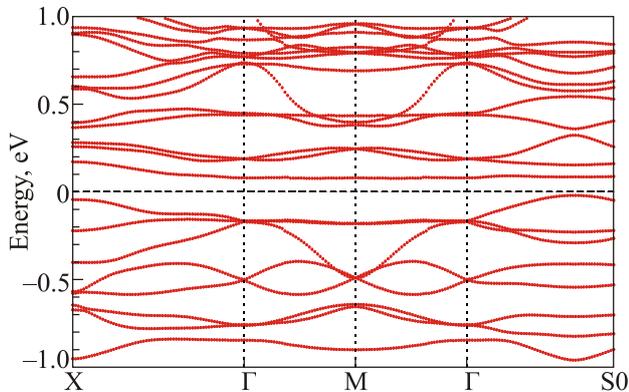

*Fig. 10.* Band structure of the insulating magnetically ordered $Rb_2Fe_4Se_5$ (245) phase in the superconducting $Rb_{0.8+x}Fe_{1.6+y}Se_2$.

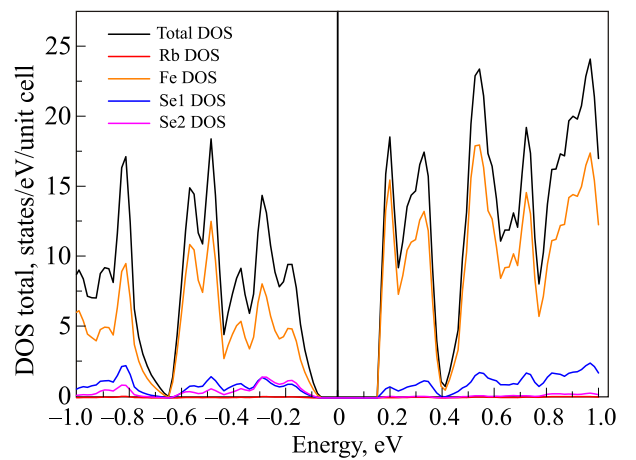

*Fig. 11.* (Color online) Density of states in the antiferromagnetic $Rb_2Fe_4Se_5$. The black line presents the total DOS. Partial DOS for Fe, Se(1), Se(2) and Rb species are shown for each muffin-tin part of the primitive cell. Note that the iron electron states contribute largely to the DOS at both side of the gap.

Table 2. Structural data of the $I4/mmm$ minority $Rb_\delta Fe_2 Se_2$ phase

| Parameter | Experiment at 200 K [35], $\delta = 0.516$ | | | Theory after optimization, $\delta = 1$; $M = 0$ | | |
|---|---|---|---|---|---|---|
| $a = b$, Å | 3.84702 | | | 3.84702 | | |
| $c$, Å | 14.7680 | | | 14.7680 | | |
| Position of atom | $x/a$ | $y/b$ | $z/c$ | $x/a$ | $y/b$ | $z/c$ |
| Rb (2a) | 0 | 0 | 0 | 0 | 0 | 0 |
| Fe (4d) | 0 | 0.5 | 0.25 | 0 | 0.5 | 0.25 |
| Se (4e) | 0 | 0 | 0.35655 | 0 | 0 | 0.331054 |

The mutual crystallography coherence of the majority and minority phases implies that a spacer should exist in between them. We suppose that the *compressed* vacancy ordered 245 phase could play a role of this spacer. Surprisingly our *ab initio* band structure calculations demonstrate that the compressed 245 phase resides in the metallic state with iron magnetic moments close to zero. The amount of this third phase can be smaller than the amount of the minority 122 phase. However, being metallic this phase may serve as the protective interface phase between the pure metallic vacancy free 122 phase and the insulating vacancy ordered 245 phase ensuring the percolative superconductivity in the $Rb_{0.77}Fe_{1.61}Se_2$.

To perform the spin-polarized band structure calculations we assume that the lattice constants of the metallic vacancy ordered $Rb_2Fe_4Se_5$ phase with $I4/m$ space symmetry are the same as the lattice constants of the minority 122 phase from [13] whis $a_{245}(met) = \sqrt{5}\, a_{122}$ and $c_{245}(met) = c_{122}$. The atomic coordinates were determined under the optimization procedure (see Table 1). We found that the $z$ coordinate of Se ions undergoes a remarkable shift and the distance between Fe–Se decreases. In addition, the difference in the Fe–Fe distances between in-block and out-of-block iron atoms which is inherent to the insulating 245 phase almost disappears in the metallic 245 phase (see Table 1).

The band structure of the compressed metallic 245 phase is shown in Fig. 13. The most prominent features of this phase are the presence of both the hole-like Fermi surface pockets at the BZ center and the electron-like Fermi surface pockets at the BZ M-point. The variety of electron bands at the Fermi surface in this phase supplies more possibilities for interband transitions than in the metallic 122 phase. Consequently, one can speculate that this spacer phase is the source of the low-intensive background. Furthermore, the coexistence of hole- and electron-like pockets at the Fermi surface in the spacer phase supports the scenario of two competing channels of the superconducting pairing in alkali iron selenides.

The frequencies and eigenvectors displacement patterns of the phonon modes were calculated using the frozen phonon approach [65] in the "fix-spin" mode for both insulating and metallic $Rb_2Fe_4Se_5$ phases. We used the struc-tural data given in Table 1. The representative displacement patterns and phonon frequencies for the Raman-active $A_g$ and $B_g$ modes in the 245 insulating phase are shown in Figs. 14 and 15.

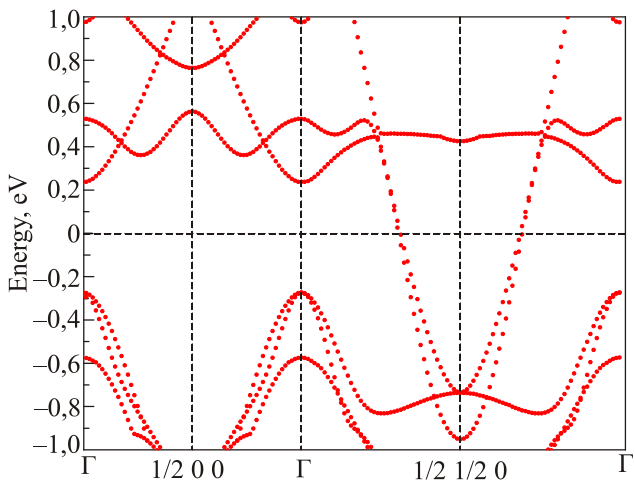

*Fig. 12*. Band structure of the metallic nonmagnetic $RbFe_2Se_2$ (122) phase in the superconducting $Rb_{0.8+x}Fe_{1.6+y}Se_2$.

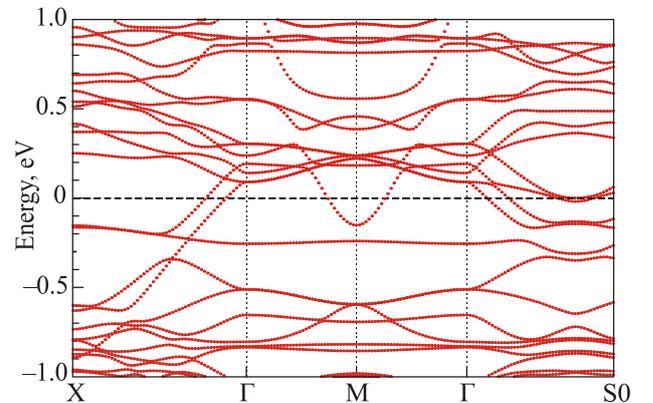

*Fig. 13*. Band structure of the metallic vacancy ordered *compressed* $Rb_2Fe_4Se_5$ phase.

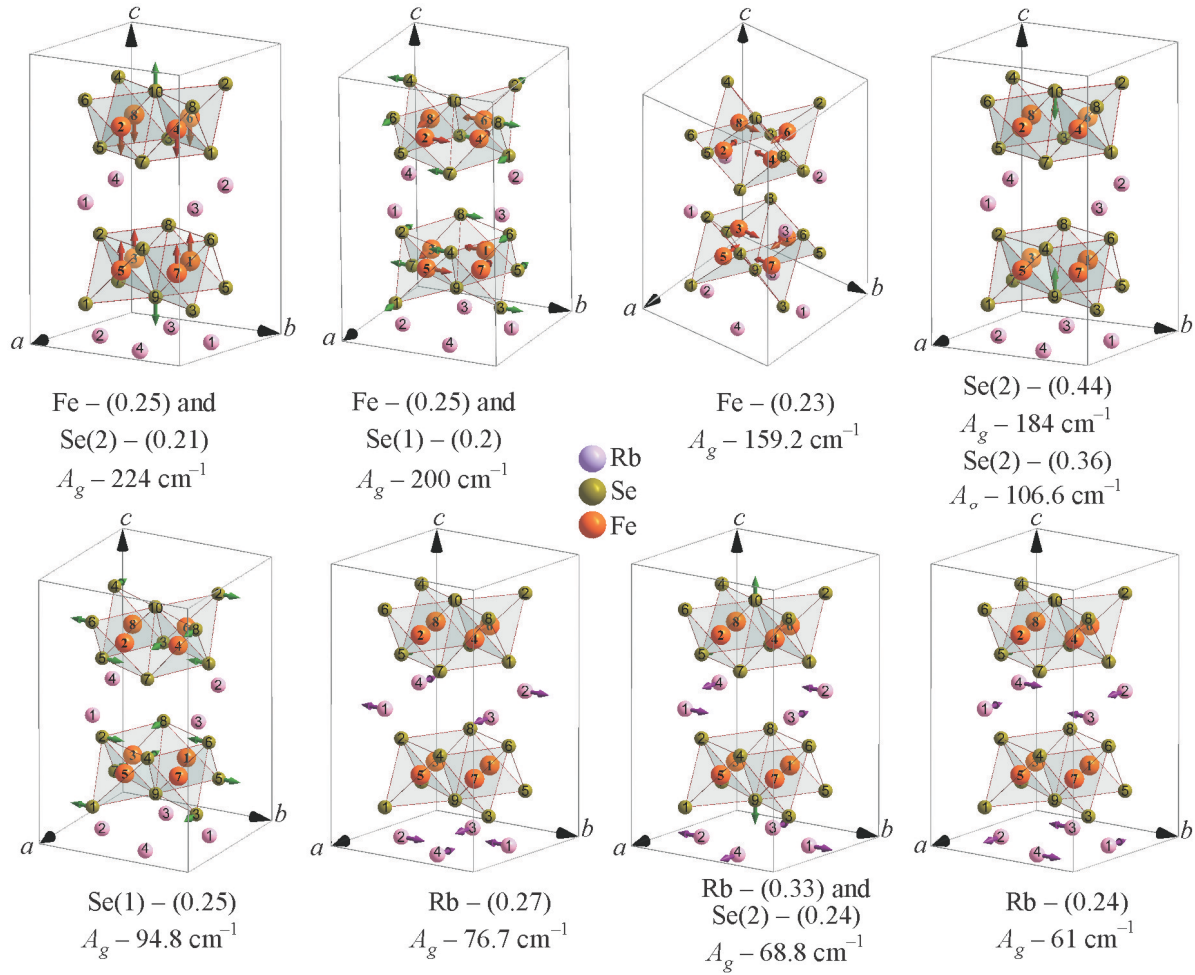

*Fig. 14.* (Color online) The calculated frequency and main atomic displacement patterns for nine Raman-active $A_g$ modes in the insulating magnetically and vacancy ordered phase of $Rb_2Fe_4Se_5$. Only those displacements exceeding 0.2 are shown with their level of displacement given in parentheses. Note the special position of the Se(2) atoms denote as 9 and 10 which are corner shared atoms of four $FeSe_4$ tetrahedrons. In accordance with the assignment shown in Table 3 three out of the four most intensive $A_g$ modes involve displacements of the Se(2) along the $z$ axis.

The phonon energies as obtained from our calculations are grouped in accordance to atomic weight of the respective vibrating ions, i.e., the high-frequency part of the phonon spectra is dominated by Fe ionic displacements (the lightest element in the composition) while the low-energy part of the phonon spectra is formed by the displacements of the most heavy Rb atoms within the almost empty $ab$-layer. This is important information for our Raman studies because such a motion implies a weak polarizability and, therefore, low-intensive Raman lines.

In Table 2 we compare experimental and theoretical data including the calculated phonon frequencies of the insulating and metallic 245 phases.

In Table 3 we denote most intensive lines in "bold" and we use "cursive" for a considerable deviation (more then 10 cm$^{-1}$) between theoretical data for the insulating 245 phase and experiment. Also we indicate by "weak" the lines which are not clearly seen at room temperature but have counterparts in our spectra at lowest temperature. In general, one can conclude a good agreement between experiment and theory for the insulating 245 phase since the difference for most frequencies does not exceed 10%. However, the theory tends to underestimate the high-frequency part spectra calculated of the insulating 245 phase and overestimates those of the conducting 245 phase.

Let's analyze the most instructive deviations between experiment and theory for the insulating phase. The largest discrepancy is found for two $A_g$ phonon modes at 106.6 and 184 cm$^{-1}$ that are mainly formed by $z$-displacements of Se(2) atoms. As was noted above, the $z$-coordinate of Se(2) atoms defines the $FeSe_4$ tetrahedral distorsions distortions and thereby affects the iron spin state. The displacements of Se(1) in the $B_g$ phonon at 177.2 cm$^{-1}$ lead to the decrease-increase of the volume of the $FeSe_4$ tetrahe-dron. This calculated phonon frequency also differs from the experimental $B_g$(4) frequency. The most striking deviations occur in the high-frequency part of the spectra, i.e., in the range of following iron atoms vibrations: the $A_g$ line at 255 cm$^{-1}$ and the $B_g$ lines at 206 and 264.4 cm$^{-1}$ at 290 K. Also, as it follows from the respective displacement patterns, these vibrations induce strong distortions of the FeSe$_4$ tetrahedron. Note that in all cases the experimental frequencies are higher than the calculated ones. Using *ab initio* lattice dynamic calculations for FeTe it was previously shown that the high-frequency part of the phonon spectrum always increases if the iron magnetic moment (the iron spin state) decreases whereas the low-lying part of the spectrum is mainly insensitive to the iron spin state [67,68]. Considering this, the calculated phonon frequencies of the metallic 245 phase are highly overestimated compared to experimental ones. While part of this deviation is related to the compressed lattice of the 245 metallic phase, it is mainly induced by the low spin state of iron.

Table 3. Comparison between experimental data and theory for $A_g$ and $B_g$ modes

| $A_g$ modes | | | $B_g$ modes | | |
|---|---|---|---|---|---|
| Experiment at 290 K | Theory: insulating 245 phase, $M = 2.62$ μ$_B$/Fe | Theory: metallic 245 phase, $M = 0$ | Experiment at 290 K | Theory: insulating 245 phase, $M = 2.62$ μ$_B$/Fe | Theory: metallic 245 phase, $M = 0$ |
| cm$^{-1}$ | | | см$^{-1}$ | | |
| $A_g$(1) – 58.7 | 61.0 | 65.8 | | 51.8 | 62.0 |
| weak | 68.8 | 72.4 | weak | 58.2 | 76.9 |
| weak | 76.7 | 85.7 | $B_g$(1) – 79.7 | 76.9 | 96.8 |
| $A_g$(2) – 94.2 | 94.8 | 141.8 | $B_g$(2) – **107.8** | 104.1 | 128.9 |
| $A_g$(3) – **120.7** | *106.6* | 154.7 | $B_g$(3) – 132.6 | 127.5 | 186.9 |
| $A_g$(4) – 155.7 | 159.3 | 208.6 | $B_g$(4) – 188.7 | *177.3* | 236.2 |
| $A_g$(5) – 173.7 | $A_{1g}$ mode of 122 phase | | $B_g$(5) – 212.4 | $B_{1g}$ mode of 122 phase | |
| $A_g$(6) – **197.6** | *184.0* | 271.7 | $B_g$(6) – 215.7 | *190.2* | 264.8 |
| $A_g$(7) – **231.5** | *199.8* | 279.1 | $B_g$(7) – **264.4** | *216.9* | 319.3 |
| $A_g$(8) – **255.7** | *224.8* | 308.1 | | | |

Thus we conclude that in the phonon calculations of the iron-based superconductors the influence of the iron-spin state is not accounted for sufficiently in the "fix-spin" mode approximation to realistically reproduce the iron involved vibrations. One should probably take a dynamical (nonadiabatic) change of the iron spin state into account which appears already during the iron atom displacements. Such a phenomenon has been observed in recent femtosecond pulse experiment of induced lattice distortions in BaFe$_2$As$_2$ where a transient change of the iron magnetic moments upon lattice oscillations has been detected [71,72].

In spite of a remarkable difference between phonon frequencies of the metallic and insulating 245 phases we do not expect to detect the former phase in the phonon Raman spectra as the fraction of the metallic 245 phase should be much smaller than the amount of the minority metallic 122 phase.

**Conclusion**

We studied the phase separation phenomenon in the superconducting single crystal Rb$_{0.77}$Fe$_{1.61}$Se$_2$ with $T_c$ = 32 K by Raman spectroscopy and *ab initio* band structure and lattice dynamics calculations. We identify phonon lines from the insulating magnetically and vacancy ordered Rb$_2$Fe$_4$Se$_5$ phase and from the vacancy free nonmagnetic and superconducting Rb$_\delta$Fe$_2$Se$_2$ phase. At temperatures below $T_c$ we observed an isotropic superconducting gap with $d_{x^2-y^2}$ symmetry. This observation agrees well with the Fermi topology which does not contain any hole pockets at the Brillouin zone center. Our band structure calculations of the 122 metallic phase confirmed this conclusion.

The only interplay between the two phases is manifested in the spectra of the insulating phase as modest anomalies in the frequency, intensity and halfwidth at the superconducting phase transition. However, some specific alteration of the electronic structure in the 122 metallic phase below $T_c$ is seen in our spectra as an enormous increase of the $B_{1g}$ (212 cm$^{-1}$) phonon intensity which belongs to this phase.

We suggest a magnetic origin for the phase separation which acts if the static magnetic moment of iron exceeds some threshold level under the magnetic ordering process. This mechanism implies the appearance of intermediate interface phases between the majority 245 insulating phase and the minority 122 metallic phase. We show that the *compressed* vacancy ordered 245 phase can be one of those intermediate phases and we highlight that this phase is metallic. This property can explain how a Josephson coupling mechanism can be realized in the superconducting M$_{0.8+x}$Fe$_{1.6+y}$Se$_2$ compounds in the presence of an insulating majority phase.

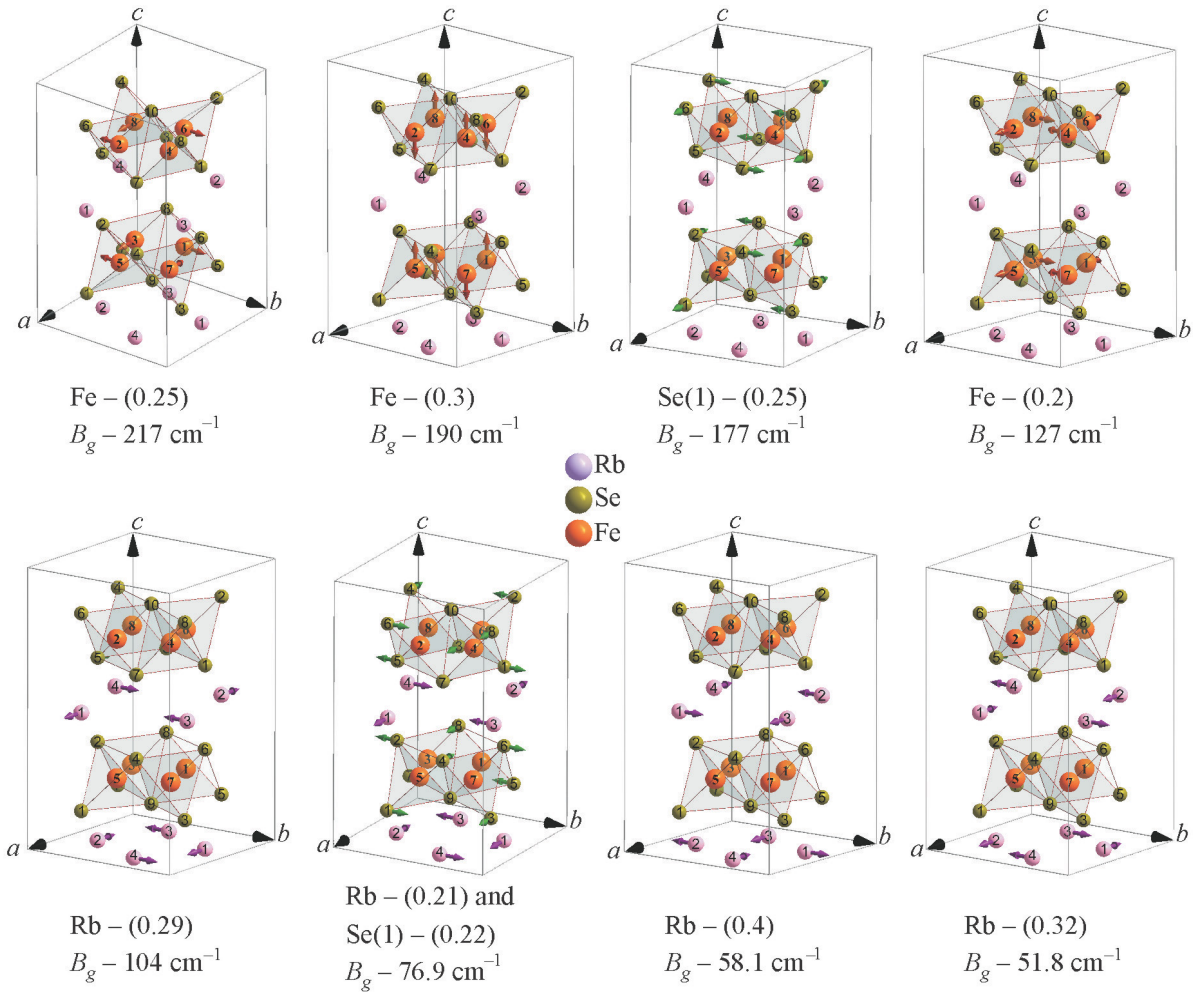

*Fig. 15.* (Color online) The calculated frequency and main atomic displacement patterns for eight Raman-active $B_g$ modes in the insulating magnetically and vacancy ordered phase of $Rb_2Fe_4Se_5$. Only those displacements exceeding 0.2 are shown with their level of displacement given in parentheses. Note that vibrations of the Se(2) atoms do not contribute to the $B_g$ modes.

Our lattice dynamics calculations show a significant difference in the phonon spectra of the conductive and insulating $Rb_2Fe_4Se_5$ phases which is mainly related to different iron spin states in these two phases. We did not observe additional lines in our spectra which could be assigned to the conductive $Rb_2Fe_4Se_5$ phase. However, we argue that the occurrence of a weak, well-structured, and highly polarized background observed in our spectra can be connected with this phase.

## Acknowledgments

*This paper is devoted to the memory of academician Kirill Borisovich Tolpygo — prominent Physicist, Teacher and Citizen, who made a great contribution to the lattice dynamics theory and many other branches of solid state physics.*

Authors thanks to Vladimir Pomjakushin for useful discussions. This work was supported in part by the State Fund of Fundamental Research of Ukraine and by the Ukrainian-Russian Grant No. 9-2010, NTH school Contacts in Nanosystems, and DFG. The calculations have been supported by resources of Ukrainian National GRID under NASU Grant No. 232. Yu. Pashkevich acknowledges partial support from the Swiss National Science Foundation (grant SNSF IZKOZ2 134161).

1. J.D. Jorgensen, B. Dabrowski, S. Pei, D.G. Hinks, L. Soderholm, B. Morosin, J.E. Schirber, E.L. Venturini, and D.S. Ginley, *Phys. Rev. B* **38**, 11337 (1988).
2. R.H. Yuan, T. Dong, Y.J. Song, G.F. Chen, J.P. Hu, J.Q. Li, and N.L. Wang, *Sci. Rep.* **2**, 221 (2012).
3. V. Ksenofontov, G. Wortmann, S. Medvedev, V. Tsurkan, J. Deisenhofer, A. Loidl, and C. Felser, *Phys. Rev. B* **84**, 180508(R) (2011).
4. A. Ricci, N. Poccia, G. Campi, B. Joseph, G. Arrighetti, L. Barba, M. Reynolds, M. Burghammer, H. Takeya, Y. Mizuguchi, Y. Takano, M. Colapietro, N.L. Saini, and A. Bianconi, *Phys. Rev. B* **84**, 060511(R) (2011).
5. A. Ricci, N. Poccia, B. Joseph, G. Arrighetti, L. Barba, J. Plaisier, G. Campi, Y. Mizuguchi, H. Takeya, Y. Takano, N.L. Saini, and A. Bianconi, *Supercond. Sci. Technol.* **24**, 082002 (2011).

6. V.Yu. Pomjakushin, E.V. Pomjakushina, A. Krzton-Maziopa, K. Conder, and Z. Shermadini, *J. Phys.: Condens. Matter* **23**, 156003 (2011).

7. Meng Wang, Miaoyin Wang, G.N. Li, Q. Huang, C.H. Li, G.T. Tan, C.L. Zhang, Huibo Cao, Wei Tian, Yang Zhao, Y.C. Chen, X.Y. Lu, Bin Sheng, H.Q. Luo, S.L. Li, M.H. Fang, J.L. Zarestky, W. Ratcliff, M.D. Lumsden, J.W. Lynn, and Pengcheng Dai, *Phys. Rev. B* **84**, 094504 (2011).

8. A. Charnukha, J. Deisenhofer, D. Pröpper, M. Schmidt, Z. Wang, Y. Goncharov, A.N. Yaresko, V. Tsurkan, B. Keimer, A. Loidl, and A.V. Boris, *Phys. Rev. B* **85** 100504 (2012).

9. Z. Shermadini, H. Luetkens, R. Khasanov, A. Krzton-Maziopa, K. Conder, E. Pomjakushina, H.-H. Klauss, and A. Amato, *Phys. Rev. B* **85**, 100501(R) (2012).

10. A. Charnukha, A. Cvitkovic, T. Prokscha, D. Pröpper, N. Ocelic, A. Suter, Z. Salman, E. Morenzoni, J. Deisenhofer, V. Tsurkan, A. Loidl, B. Keimer, and A.V. Boris, *Phys. Rev. Lett.* **109**, 017003 (2012).

11. Y. Texier, J. Deisenhofer, V. Tsurkan, A. Loidl, D.S. Inosov, G. Friemel, and J. Bobroff, *Phys. Rev. Lett.* **108**, 237002 (2012).

12. J. Maletz, V.B. Zabolotnyy, D.V. Evtushinsky, A.N. Yaresko, A.A. Kordyuk, Z. Shermadini, H. Luetkens, K. Sedlak, R. Khasanov, A. Amato, A. Krzton-Maziopa, K. Conder, E. Pomjakushina, H.-H. Klauss, E.D.L. Rienks, B. Büchner, and S.V. Borisenko, *Phys. Rev. B* **88**, 134501 (2013).

13. V.Yu. Pomjakushin, A. Krzton-Maziopa, E.V. Pomjakushina, K. Conder, D. Chernyshov, V. Svitlyk, and A. Bosak, *J. Phys.: Condens. Matter* **24**, 435701 (2012).

14. A. Bosak, V. Svitlyk, A. Krzton-Maziopa, E. Pomjakushina, K. Conder, V. Pomjakushin, A. Popov, D. de Sanctis, D. Chernyshov, *Phys. Rev. B* **86**, 174107 (2012).

15. V. Svitlyk, D. Chernyshov, A. Bosak, E. Pomjakushina, A. Krzton-Maziopa, K. Conder, V. Pomjakushin, V. Dmitriev, G. Garbarino, and M. Mezouar, *Phys. Rev. B* **89**, 144106 (2014).

16. Y. Liu, Q. Xing, W.E. Straszheim, J. Marshman, P. Pedersen, R. McLaughlin, and T.A. Lograsso, *Phys. Rev. B* **93**, 064509 (2016).

17. A. Ricci, N. Poccia, B. Joseph, D. Innocenti, G. Campi, A. Zozulya, F. Westermeier, A. Schavkan, F. Coneri, A. Bianconi, H. Takeya, Y. Mizuguchi, Y. Takano, T. Mizokawa, M. Sprung, and N.L. Saini, *Phys. Rev. B* **91**, 020503 (2015).

18. Rong Yu and Qimiao Si, *Phys. Rev. Lett.* **110**, 146402 (2013).

19. M. Yi, D. Lu, R. Yu, S. Riggs, J.-H. Chu, B. Lv, Z. Liu, M. Lu, Y.-T. Cui, M. Hashimoto, S.-K. Mo, Z. Hussain, C. Chu, I. Fisher, Q. Si, and Z.-X. Shen, *Phys. Rev. Lett.* **110**, 067003 (2013).

20. Z. Wang, M. Schmidt, J. Fischer, V. Tsurkan, M. Greger, D. Vollhardt, A. Loidl, and J. Deisenhofer, *Nat. Commun.* **5**, 3202 (2014).

21. M.Yi, Z.-K. Liu, Y. Zhang, R. Yu, J.-X. Zhu, J.J. Lee, R.G. Moore, F.T. Schmitt, W. Li, S.C. Riggs, J.-H. Chu, B. Lv, J. Hu, M. Hashimoto, S.-K. Mo, Z. Hussain, Z.Q. Mao, C.W. Chu, I.R. Fisher, Q. Si, Z.-X. Shen, and D.H. Lu, *Nat. Commun*. **6**, 7777 (2015).

22. N. Lazarević, M. Abeykoon, P.W. Stephens, Hechang Lei, E.S. Bozin, C. Petrovic, and Z.V. Popović, *Phys. Rev. B* **86** 054503 (2012).

23. S. Tsuda, Y. Mizuguchi, H. Takeya, T. Yamaguchi, and Y. Takano, *J. Phys. Soc. Jpn*. **80**, 075003 (2011).

24. N. Lazarević, Hechang Lei, C. Petrovic, and Z.V. Popović, *Phys. Rev. B* **84**, 214305 (2011).

25. A.M. Zhang, K. Liu, J.H. Xiao, J.B. He, D.M. Wang, G.F. Chen, B. Normand, and Q.M. Zhang, *Phys. Rev. B* **85**, 024518 (2012).

26. A.M. Zhang, K. Liu, J.B. He, D.M. Wang, G.F. Chen, B. Normand, and Q.M. Zhang, *Phys. Rev. B* **86**, 134502 (2012).

27. A. Ignatov, A. Kumar, P. Lubik, R.H. Yuan, W.T. Guo, N.L. Wang, K. Rabe, and G. Blumberg, *Phys. Rev. B* **86**, 134107 (2012).

28. F. Kretzschmar, B. Muschler, T. Böhm, A. Baum, R. Hackl, Hai-Hu Wen, V. Tsurkan, J. Deisenhofer, and A. Loidl, *Phys. Rev. Lett.* **110**, 187002 (2013).

29. F. Ye, S. Chi, Wei Bao, X.F. Wang, J.J. Ying, X.H. Chen, H.D. Wang, C.H. Dong, and Minghu Fang, *Phys. Rev. Lett.* **107**, 137003 (2011).

30. S. Weyeneth, M. Bendele, F. von Rohr, P. Dluzewski, R. Puzniak, A. Krzton-Maziopa, S. Bosma, Z. Guguchia, R. Khasanov, Z. Shermadini, A. Amato, E. Pomjakushina, K. Conder, A. Schilling, and H. Keller, *Phys. Rev. B* **86**, 134530 (2012).

31. S.C. Speller, T.B. Britton, G.M. Hughes, A. Krzton-Maziopa, E. Pomjakushina, K. Conder, A.T. Boothroyd, and C.R.M. Grovenor, *Supercond. Sci. Technol*. **25**, 084023 (2012).

32. S.C. Speller, P. Dudin, S. Fitzgerald, G.M. Hughes, K. Kruska, T.B. Britton, A. Krzton-Maziopa, E. Pomjakushina, K. Conder, A. Barinov, and C.R.M. Grovenor, *Phys. Rev. B* **90**, 024520 (2014).

33. C.C. Homes, Z.J. Xu, J.S. Wen, and G.D. Gu, *Phys. Rev. B* **85**, 180510(R) (2012).

34. Scott V. Carr, Despina Louca, Joan Siewenie, Q. Huang, Aifeng Wang, Xianhui Chen, and Pengcheng Dai, *Phys. Rev. B* **89**, 134509 (2014).

35. Y. Kobayashi, S. Kototani, K. Ohishi, M. Itoh, A. Hoshikawa, T. Ishigaki, and M. Sato, *J. Phys. Soc. Jpn.* **84**, 044710 (2015).

36. F. Chen, M. Xu, Q.Q. Ge, Y. Zhang, Z.R. Ye, L.X. Yang, Juan Jiang, B.P. Xie, R.C. Che, M. Zhang, A.F. Wang, X.H. Chen, D.W. Shen, X.M. Xie, M.H. Jiang, J.P. Hu, and D.L. Feng, *Phys. Rev. X* **1**, 021020 (2011).

37. Y. Zhang, L.X. Yang, M. Xu, Z.R. Ye, F. Chen, C. He, J. Jiang, B.P. Xie, J.J. Ying, X.F. Wang, X.H. Chen, J.P. Hu, and D.L. Feng, *Nature Mater.* **10**, 273 (2011).

38. T. Qian, X.-P. Wang, W.-C. Jin, P. Zhang, P. Richard, G. Xu, X. Dai, Z. Fang, J.-G. Guo, X.-L. Chen, and H. Ding, *Phys. Rev. Lett.* **106**, 187001 (2011).

39. L. Zhao, D. Mou, S. Liu, X. Jia, J. He, Y. Peng, L. Yu, X. Liu, G. Liu, S. He, X. Dong, J. Zhang, J.B. He, D.M. Wang, G.F.

Chen, J.G. Guo, X.L. Chen, X. Wang, Q. Peng, Z. Wang, S. Zhang, F. Yang, Z. Xu, C. Chen, and X.J. Zhou, *Phys. Rev. B* **83**, 140508(R) (2011).

40. D. Mou, S. Liu, X. Jia, J. He, Y. Peng, L. Zhao, L. Yu, G. Liu, S. He, X. Dong, J. Zhang, H. Wang, C. Dong, M. Fang, X. Wang, Q. Peng, Z. Wang, S. Zhang, F. Yang, Z. Xu, C. Chen, and X.J. Zhou, *Phys. Rev. Lett.* **106**, 107001 (2011).
41. E. Dagotto, *Rev. Mod. Phys.* **85**, 849 (2013).
42. V. Tsurkan, J. Deisenhofer, A. Günther, H.-A. Krug von Nidda, S. Widmann, and A. Loidl, *Phys. Rev. B* **84**, 144520 (2011).
43. A. Krzton-Maziopa, E. Pomjakushina, and K. Conder, *J. Crystal Growth* **360**, 155 (2012).
44. I.A. Nekrasov and M.V. Sadovskii, *JETP Lett.* **93**, 166 (2011).
45. V.P. Gnezdilov, Yu.G. Pashkevich, J.M. Tranquada, P. Lemmens, G. Güntherodt, A.V. Yeremenko, S.N. Barilo, S.V. Shiryaev, L.A. Kurnevich, and P.M. Gehring, *Phys. Rev. B* **69**, 174508 (2004).
46. A. Krzton-Maziopa, Z. Shermadini, E. Pomjakushina, V. Pomjakushin, M. Bendele, A. Amato, R. Khasanov, H. Luetkens, and K. Conder, *J. Phys.: Condens. Matter* **23**, 052203 (2011).
47. V. Gnezdilov, Yu.G. Pashkevich, P. Lemmens, D. Wulferding, T. Shevtsova, A. Gusev, D. Chareev, and A. Vasiliev, *Phys. Rev. B* **87**, 144508 (2013).
48. M. Wang, C. Li, D.L. Abernathy, Y. Song, S.V. Carr, X. Lu, S. Li, Z. Yamani, J. Hu, T. Xiang, and P. Dai, *Phys. Rev. B* **86**, 024502 (2012).
49. S. Chi, F.Ye, W.Bao, M. Fang, H. Wang, C. Dong, A. Savici, G. Granroth, M. Stone, and R. Fishman, *Phys. Rev. B* **87**, 100501 (2013).
50. B. Valenzuela, M.J. Calderon, G. León, and E. Bascones, *Phys. Rev. B* **87**, 075136 (2013).
51. V.K. Thorsmølle, M. Khodas, Z.P. Yin, Chenglin Zhang, S.V. Carr, Pengcheng Dai, and G. Blumberg, *Phys. Rev. B* **93**, 054515 (2016).
52. I.J. Pomeranchuk, *Sov. Phys. JETP* **8**, 361 (1959).
53. T. Devereaux and R. Hackl, *Rev. Mod. Phys.* **79**, 175 (2007).
54. A. Kreisel, Y. Wang, T.A. Maier, P.J. Hirschfeld, and D.J. Scalapino, *Phys. Rev. B* **88**, 094522 (2013).
55. A. Bardasis and J.R. Schrieffer, *Phys. Rev.* **121**, 1050 (1961).
56. D.J. Scalapino and T.P. Devereaux, *Phys. Rev. B* **80**, 140512 (2009).
57. M. Khodas, A.V. Chubukov, and G. Blumberg, *Phys. Rev. B* **89**, 245134 (2014).
58. Bo Li, Lihua Pan, Yuan-Yen Tai, Matthias J. Graf, Jian-Xin Zhu, Kevin E. Bassler, and C.S. Ting, *Phys. Rev. B* **91** 220509 (2015).
59. M. Bendele, A. Amato, K. Conder, M. Elender, H. Keller, H.-H. Klauss, H. Luetkens, E. Pomjakushina, A. Raselli, and R. Khasanov, *Phys. Rev. Lett.* **104**, 087003 (2010).
60. Z.G. Chen, R.H. Yuan, T. Dong, G. Xu, Y.G. Shi, P. Zheng, J.L. Luo, J.G. Guo, X.L. Chen, and N.L. Wang, *Phys. Rev. B* **83**, 220507(R) (2011).
61. I.I. Mazin, M.D. Johannes, L. Boeri, K. Koepernik, and D.J. Singh, *Phys. Rev. B* **78**, 085104 (2008).
62. T. Yildirim, *Phys. Rev. Lett.* **102**, 037003 (2009); *Physica C* **469**, 425 (2009).
63. M. Zbiri, H. Schober, M.R. Johnson, S. Rols, R. Mittal, Y. Su, M. Rotter, and D. Johrendt, *Phys. Rev. B* **79**, 064511 (2009).
64. D. Reznik, K. Lokshin, D.C. Mitchell, D. Parshall, W. Dmowski, D. Lamago, R. Heid, K.-P. Bohnen, A.S. Sefat, M.A. McGuire, B.C. Sales, D.G. Mandrus, A. Subedi, D.J. Singh, A. Alatas, M.H. Upton, A.H. Said, A. Cunsolo, Yu. Shvyd'ko, and T. Egam, *Phys. Rev. B* **80**, 214534 (2009).
65. http://elk.sourceforge.net
66. John P. Perdew and Yue Wang, *Phys. Rev. B* **45**, 13244 (1992); D.M. Ceperley and B.J. Alder, *Phys. Rev. Lett.* **45**, 566 (1980); J.P. Perdew, A. Ruzsinszky, G.I. Csonka, O.A. Vydrov, G.E. Scuseria, L.A. Constantin, Z. Zhou, and K. Burke, *Phys. Rev. Lett.* **100**, 136406 (2008).
67. V. Gnezdilov, Yu. Pashkevich, P. Lemmens, A. Gusev, K. Lamonova, T. Shevtsova, I. Vitebskiy, O. Afanasiev, S. Gnatchenko, V. Tsurkan, J. Deisenhofer, and A. Loidl, *Phys. Rev. B* **83**, 245127 (2011).
68. Yu.G. Pashkevich, T.N. Shevtsova, A.A. Gusev, V.P. Gnezdilov, and P. Lemmens, *Fiz. Nizk. Temp.* **38**, (2012) [*Low Temp. Phys.* **38**, 900 (2012)].
69. X.-W. Yan, M. Gao, Z.-Y. Lu, and T. Xiang, *Phys. Rev. B* **83**, 233205 (2011).
70. C. Cao and J. Dai, *Phys. Rev. Lett.* **107**, 056401 (2011).
71. L. Rettig, S.O. Mariager, A. Ferrer, S. Grübel, J.A. Johnson, J. Rittmann, T. Wolf, S.L. Johnson, G. Ingold, P. Beaud, and U. Staub, *Phys. Rev. Lett.* **114,** 067402 (2015).
72. S. Gerber, K.W. Kim, Y. Zhang, D. Zhu, N. Plonka, M.Yi, G.L. Dakovski, D. Leuenberger, P.S. Kirchmann, R.G. Moore, M. Chollet, J.M. Glownia, Y. Feng, J.-S. Lee, A. Mehta, A.F. Kemper, T. Wolf, Y.-D. Chuang, Z. Hussain, C.-C. Kao, B. Moritz, Z.-X. Shen, T.P. Devereaux, and W.-S. Lee, *Nat. Commun.* **6**, 7377 (2015).